\documentclass{aastex63}

\usepackage{setspace}

\usepackage{subfigure}
\usepackage{comment}

\begin{document}

\newcommand{\average}[1]{\ensuremath{\langle#1\rangle} }

\title{Impact of hypernova $\nu p$-process nucleosynthesis on the galactic chemical evolution of Mo and Ru}

\author{Hirokazu Sasaki}
\affiliation{Division of Science, National Astronomical Observatory of Japan, \\
2-21-1 Osawa, Mitaka, Tokyo 181-8588, Japan}

\author{Yuta Yamazaki}
\affiliation{Division of Science, National Astronomical Observatory of Japan, \\
2-21-1 Osawa, Mitaka, Tokyo 181-8588, Japan}
\affiliation{Graduate School of Science, The University of Tokyo, \\
7-3-1 Hongo, Bunkyo-ku, Tokyo 113-033, Japan}

\author{Toshitaka Kajino}
\affiliation{Division of Science, National Astronomical Observatory of Japan, \\
2-21-1 Osawa, Mitaka, Tokyo 181-8588, Japan}
\affiliation{Graduate School of Science, The University of Tokyo, \\
7-3-1 Hongo, Bunkyo-ku, Tokyo 113-033, Japan}
\affiliation{School of Physics, and International Research Center for Big-Bang Cosmology and Element Genesis, Beihang University,\\
37 Xueyuan Rd., Haidian-district, Beijing 100083 China}

\author{Motohiko Kusakabe}
\affiliation{School of Physics, and International Research Center for Big-Bang Cosmology and Element Genesis, Beihang University,\\
37 Xueyuan Rd., Haidian-district, Beijing 100083 China}

\author{Takehito Hayakawa}
\affiliation{National Institutes for Quantum and Radiological Science and Technology, \\
2-4 Shirakata, Tokai, Naka, Ibaraki 319-1106, Japan}
\affiliation{Institute of Laser Engineering, Osaka University, Suita, Osaka 565-0871, Japan}

\author{Myung-Ki Cheoun}
\affiliation{Department of Physics and OMEG Institute, Soongsil University, Seoul 156-743, Korea}

\author{Heamin Ko}
\affiliation{Department of Physics and OMEG Institute, Soongsil University, Seoul 156-743, Korea}

\author{Grant J. Mathews}
\affiliation{Department of Physics and Center for Astrophysics, University of Notre Dame, Notre Dame, IN 46556, USA}











\setstretch{1.5}
\large
\begin{abstract}


We calculate the Galactic Chemical Evolution (GCE) of Mo and Ru by taking into account the contribution from $\nu p$-process nucleosynthesis. We estimate yields of $p$-nuclei such as $^{92,94}\mathrm{Mo}$ and $^{96,98}\mathrm{Ru}$ through the $\nu p$-process in various supernova (SN) progenitors based upon recent models. In particular, the $\nu p$-process in energetic  hypernovae produces a large amount of $p$-nuclei compared to the yield in ordinary core-collapse SNe. Because of this the abundances of $^{92,94}\mathrm{Mo}$ and $^{96,98}\mathrm{Ru}$ in the Galaxy are significantly enhanced at [Fe/H]=0 by the $\nu p$-process. We find that the $\nu p$-process in hypernovae is the main contributor to the elemental abundance of $^{92}$Mo at low metallicity [Fe/H$]<-2$. Our theoretical prediction of the elemental abundances in metal-poor stars becomes more consistent with observational data when the $\nu p$-process in hypernovae is taken into account.

\end{abstract}

\section{Introduction} 
\label{sec:introduction}

The $p$-nuclei are stable nuclides with atomic numbers $Z \geq 34$ that are located on the proton-rich side of the $\beta$-stability line ~\citep{Burbidge1957}. The isotopic fractions of the $p$-nuclei are typically lower than 1.5\%. Such $p$-nuclei are not synthesized through the rapid or slow neutron capture processes ($r$- and $s$-processes) that predominantly contribute to production of heavy nuclei.  It was suggested that the $p$-nuclei may be produced by ($p$,~$\gamma$) or ($\gamma$,~$p$) reactions \citep{Burbidge1957}. \citet{Wooley1978} and \citet{Hayakawa2004} have provided evidence confirming that most $p$-nuclei are synthesized by successive photodisintegration reactions from heavier isotopes in high temperature environments ($\gamma$-process). 

Candidates for the $\gamma$-process astrophysical site include the  O-Ne rich layers in core-collapse supernovae (SNe II)~\citep{Wooley1978,Rayet1990,Prantzos1990,Rayet1995,Hayakawa2008} and the outermost layers of an exploding carbon-oxygen white dwarf (SNe Ia)~\citep{Howard1991,Goriely2002,Kusakabe2011,Travaglio2011}.  

Calculations of $\gamma$-process nucleosynthesis can produce the relative abundances of most $p$-nuclei except for $^{92,94}$Mo and $^{96,98}$Ru.
The solar isotopic fractions of these four nuclei are $14.84$, $9.25$, $5.54$, and $1.87$\%, respectively.  These fractions are much higher than those of other $p$-nuclei.
The relatively large isotopic fractions suggest another major astrophysical process
for the origin of these four nuclei.
\citet{Howard1991} calculated the nucleosynthesis in SNe Ia to explain the abundances of the Mo and Ru isotopes by a  combination of photodisintegration reactions and particle-induced reactions.
However, the results of Galactic Chemical Evolution (GCE) simulations with SNe Ia \citep{Travaglio2015} and SNe II \citep{Travaglio2018} underproduce the solar abundances of the Mo and Ru $p$-isotopes.

\citet{Frohlich2006} proposed a new mechanism,  $\nu p$-process nucleosynthesis.  This process occurs via free neutrons produced from the absorption of electron antineutrinos on free protons, $p(\bar{\nu}_{e},~e^{+})n$, in the proton-rich neutrino-driven winds of SNe II. This enhances the abundances of $^{92,94}\mathrm{Mo}$ and $^{96,98}\mathrm{Ru}$. Large amounts of $p$-nuclei can be produced through a series of ($p$,~$\gamma$) and ($n$,~$p$) reactions in the $\nu p$-process \citep{Frohlich2006,Pruet2006,Wanajo2006}. The uncertainties in the $\nu p$-process such as the hydrodynamic state variables of neutrino-driven winds, neutrino fluxes, and nuclear reaction rates have been studied widely  \citep{Wanajo2011,Arcones2012,Fujibayashi2015,Sasaki2017,Bliss2018,Nishimura2019,Xiong2020,Jin2020}.
However, there has been little  clear observational evidence, for example in solar abundances, stellar chemical composition, or isotopic anomalies in primitive meteorites, that these proposed processes actually occurred in the Galaxy.  

Recently, however, it was reported that the abundance ratios observed in the Cassiopeia A SN remnant are results of the $\alpha$-rich freeze out and neutrino-processed proton rich ejecta \citep{Sato2021}. 
Furthermore, molybdenum isotopic anomalies in both differentiated (i.e. melted asteroids)  and primitive meteorites (i.e. carbonaceous chondrites) have been reported \citep{Dauphas2002,Budde2016,Poole2017}. In particular it was found that the enhanced isotopic anomalies of the $p$-isotopes $^{92,94}$Mo are correlated with that of the $r$-isotope $^{100}$Mo \citep{Dauphas2002,Budde2016}.
However, isotopic abundances observed in iron  meteorites \citep{Poole2017} shows a different pattern. In this case, the anomaly of $^{92}$Mo only weakly correlates with that of $^{100}$Mo. These results indicate that $^{92,94}$Mo may be synthesized in the same star that produces $^{100}$Mo but by a different nucleosynthetic process.

The ${\nu}p$-process in core-collapse SNe where the $r$-process also occurs is one of the candidate sites for the production of $^{92,94}$Mo and $^{96,98}$Ru.
The observational evidence that these four nuclei are predominately synthesized by the $\nu p$-process may be found in metal-poor stars whose Mo and Ru elemental compositions have been reported.

For the present  work we study the GCE of Mo and Ru.  We first determine yield data of the $\nu p$-process in various SN models.  A range of models is considered because these processes depend upon the hydrodynamic state variables of neutrino-driven winds in SNe II.  In particular, we estimate yields of $^{92,94}\mathrm{Mo}$ and $^{96,98}\mathrm{Ru}$ in the $\nu p$-process based upon simulation results of both SNe II and hypernovae (HNe) that are more energetic SN events ($\sim10^{52}$erg) \citep{Galama1998,Iwamoto1998}.
We then apply the calculated yields of these four anomalous $p$-nuclei to GCE and analyze  the various contributions from the $\nu p$-process to their solar abundances.
We show that the yield of the $\nu p$-process in hypernovae can explain the enhanced  elemental abundances of Mo and Ru observed in metal-poor stars.

We note that a complementary study of the GCE for Mo and Ru (along with Sr, Y, Zr, and Ba) has recently been reported in \citet{Vincenzo2021}.  In that work it was similarly concluded that the production of Mo and Ru in proto-neutron star (PNS) winds from SNe are insufficient to account for the observed abundances of Mo and Ru at low metallicity without arbitrarily enhancing the production by about a factor of  30.  They then considered that the required enhanced production factor might result from effects of a rapidly rotating PNS (although these models overpredict  [Sr/Fe] and [Mo/Fe] at higher metallicity).  The present  work differs from that study in that we adopt standard production factors, and demonstrate that the required additional production at low metallicity can be attributed to the $\nu p$-process contributions from HNe.

\section{$\nu$\lowercase{p}-process}



We estimate the abundances of $p$-nuclei produced through the $\nu p$-process in both SNe II and HNe. In the case of SNe II, the production of $p$-nuclei by the $\nu p$-process is small in the early explosion phase ($t<1$s) because of the low entropy per baryon and long timescale associated with the early dynamical ejecta \citep{Wanajo2018}. 
We therefore focus on the $\nu p$-process in the later explosion phase ($t>1$s). We construct general relativistic steady-state, spherically symmetric neutrino driven winds as described in \citet{Otsuki2000}. These models are calibrated with numerical results of $9,12,25$, and $60M_{\odot}$ progenitor models in recent $3$D core-collapse SN simulations \citep{Burrows2020,Nagakura2021}. The hydrodynamic quantities associated with the  wind trajectories for each different progenitor model are derived by evolving the electron fraction $Y_{e}$, the neutrino luminosity $L_{\nu}$, the neutrino mean energy $\average{E_{\nu}}$, the radius of the PNS $R_{\nu}$, and the gravitational mass of the PNS $M_{\mathrm{PNS}}$. 

Existing hydrodynamic simulations of  $3$D explosion models stop at $\sim1$ s. 
We extrapolate them to later times by
assuming an exponential decrease in the neutrino luminosity after the hydrodynamic simulations end  at $t=t_{0}$, $L_{\nu}(t)=L_{\nu,0}\exp(-(t-t_{0})/\tau_{\nu})$, as used in \citet{Woosley1990}.  Here, $L_{\nu,0}$ is the neutrino luminosity at $t=t_{0}$ and $\tau_{\nu}=1$ s is the timescale for neutrino cooling. The values of $t_{0}$ in the different models are shown in Table $2$ of \citet{Burrows2020}. We also assume that the evolution of the PNS radius is given by $R_{\nu}(L_{\nu})=(R_{0}-R_{1})(L_{\nu}/L_{\nu,0})+R_{1}$ \citep{Wanajo2013} where $R_{0}$ represents the radius of the PNS at $t=t_{0}$ and $R_{1}$ is the radius of the PNS after the neutrino cooling. We set $R_{1}=12$ km, which satisfies the observational constraints on the radii of cold neutron stars \citep{Steiner2013}. 
Furthermore, we take into account the decrease of the neutrino temperature as the neutrino cooling proceeds. 
We assume that the neutrinos obey Fermi-Dirac distributions without chemical potentials, where the neutrino temperature $T_{\nu}=\average{E_{\nu}}/3.151$ is proportional to $(L_{\nu}R_{\nu}^{-2})^{1/4}$ when $t>t_{0}$ \citep{Balantekin2005}. 
This relation induces a decrease of $T_{\nu}$ as $L_{\nu}$ is reduced. Nucleosynthesis in the neutrino driven wind is calculated using nuclear network simulations. We follow the same numerical setup of the network simulations as that of \citet{Sasaki2017}. We ignore the effect of neutrino oscillations because the energy spectra of different neutrino species in the present models are almost degenerate in the later explosion phase ($t>1$ s). The total yield of a nucleus $i$ inside a neutrino driven wind is written as $y_{i}=\int_{t=1\mathrm{s}}\mathrm{d}t\ X_{i}(t)\dot{M}(t)$ where $\dot{M}(t)$ and $X_{i}(t)$ are the mass ejection rate and the mass fraction of the nucleus $i$ inside the wind trajectory at time $t$. The total yields of $p$-nuclei in neutrino driven winds for the progenitor models with the initial masses of $9,12,25$, and $60M_{\odot}$ are used as input data for the subsequent GCE calculations. 

The $\nu p$-process in the HN model is based upon neutrino-driven winds obtained in \citet{Fujibayashi2015} where the possible synthesis of heavy elements in a $100M_{\odot}$ progenitor star was investigated. In this progenitor model, a massive PNS ($\sim3M_{\odot}$) is maintained for a few seconds before the black hole forms.
We adopt the neutrino-driven wind model (e) in Table $3$ of \citet{Fujibayashi2015} as the fiducial model of proton-rich neutrino-driven winds in HNe. The yield of  $p$-nuclei is obtained by multiplying a typical lifetime of massive PNSs ($\tau_{NS} \approx 1$ s) by the mass ejection rate ($\dot{M}$) and the mass fraction ($X_{i}$) of the $p$-nucleus in the trajectory of model (e).

\begin{figure}
\subfigure{%
    \includegraphics[clip, width=0.5\columnwidth]{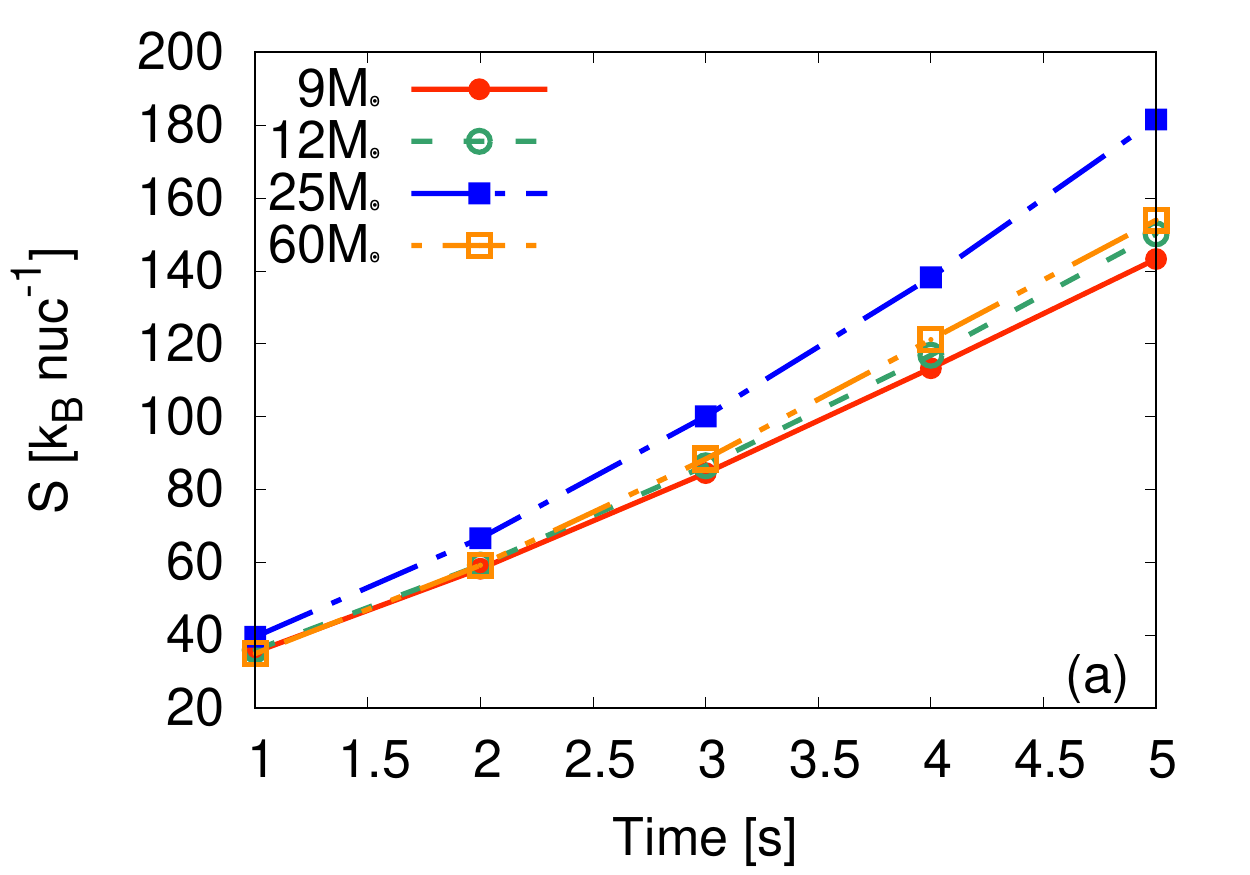}}%
\subfigure{%
    \includegraphics[clip, width=0.5\columnwidth]{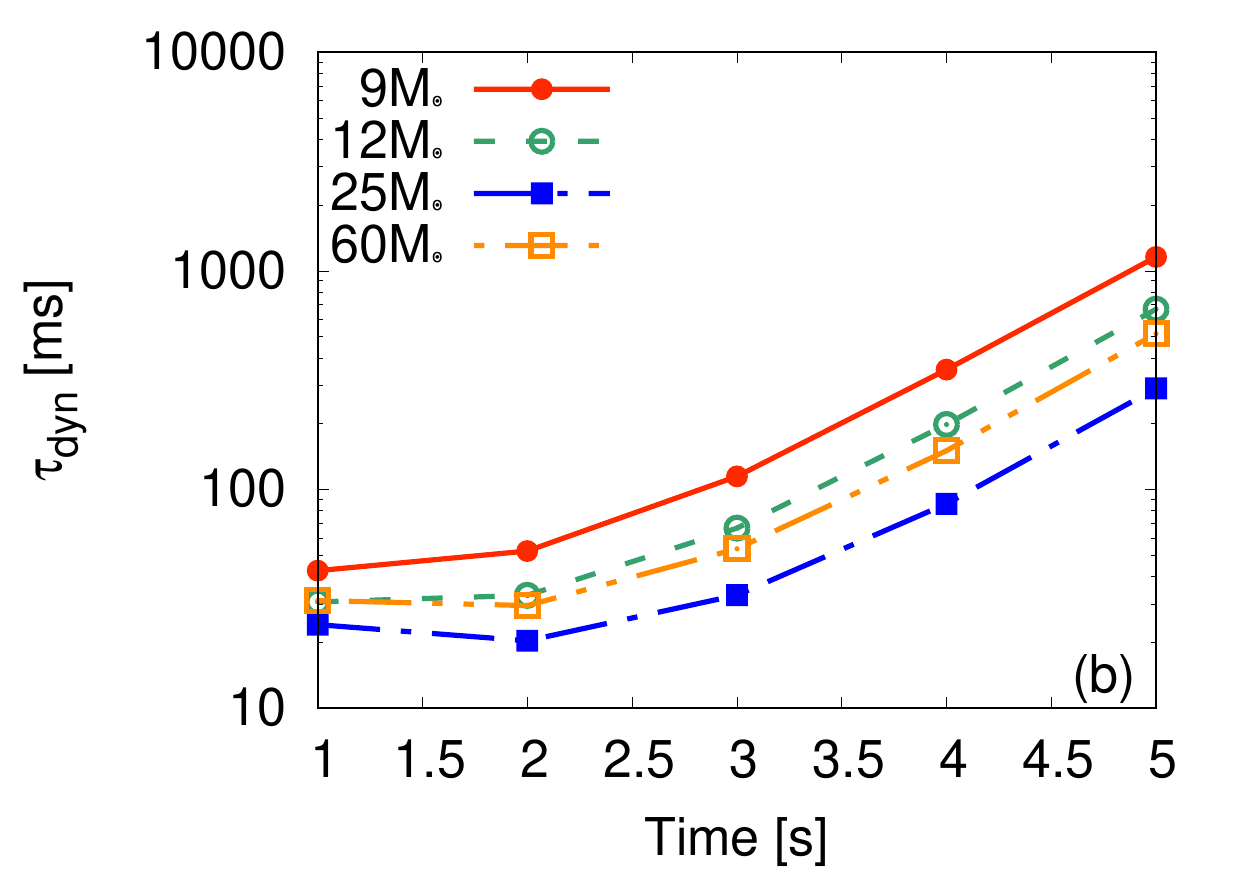}}%
\end{figure}
\begin{figure}
\subfigure{%
    \includegraphics[clip, width=0.5\columnwidth]{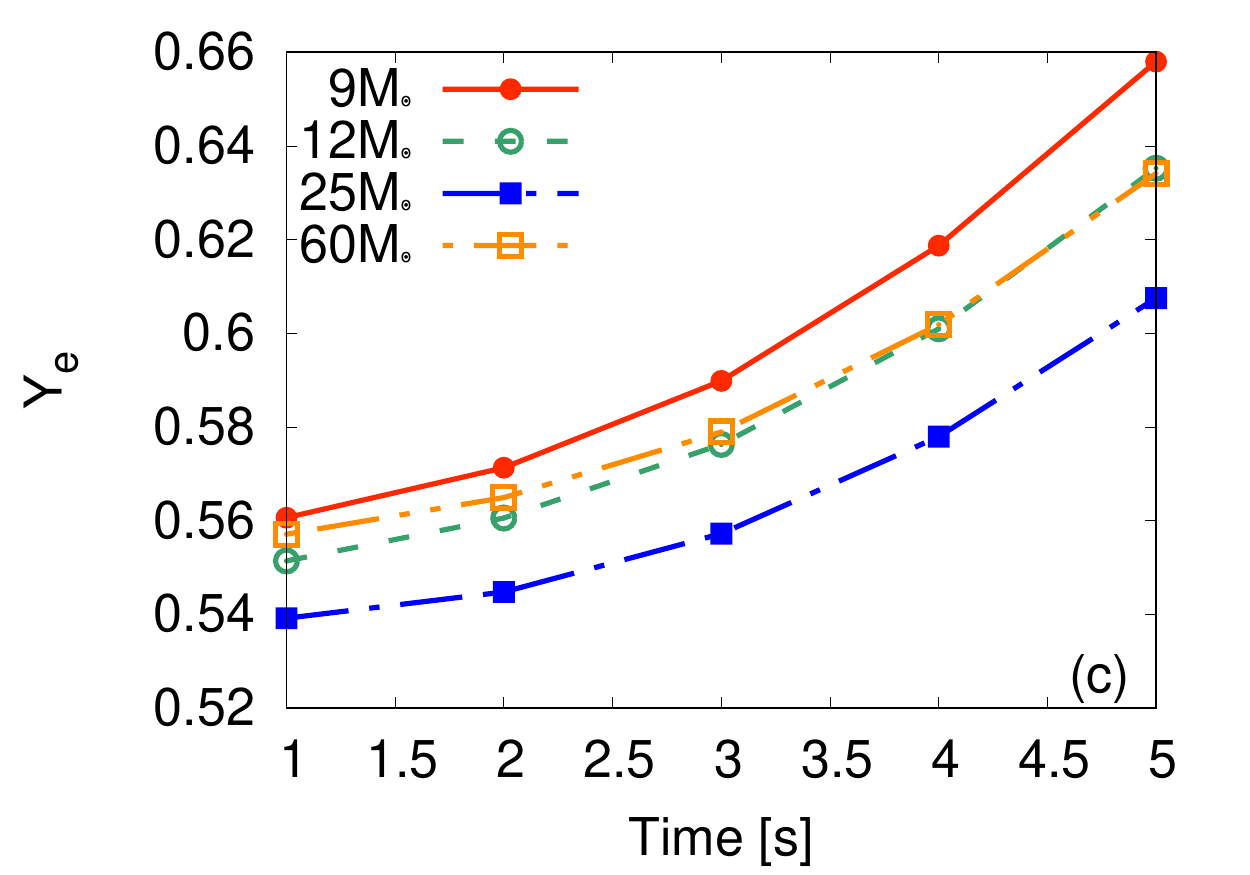}}%
\subfigure{%
    \includegraphics[clip, width=0.5\columnwidth]{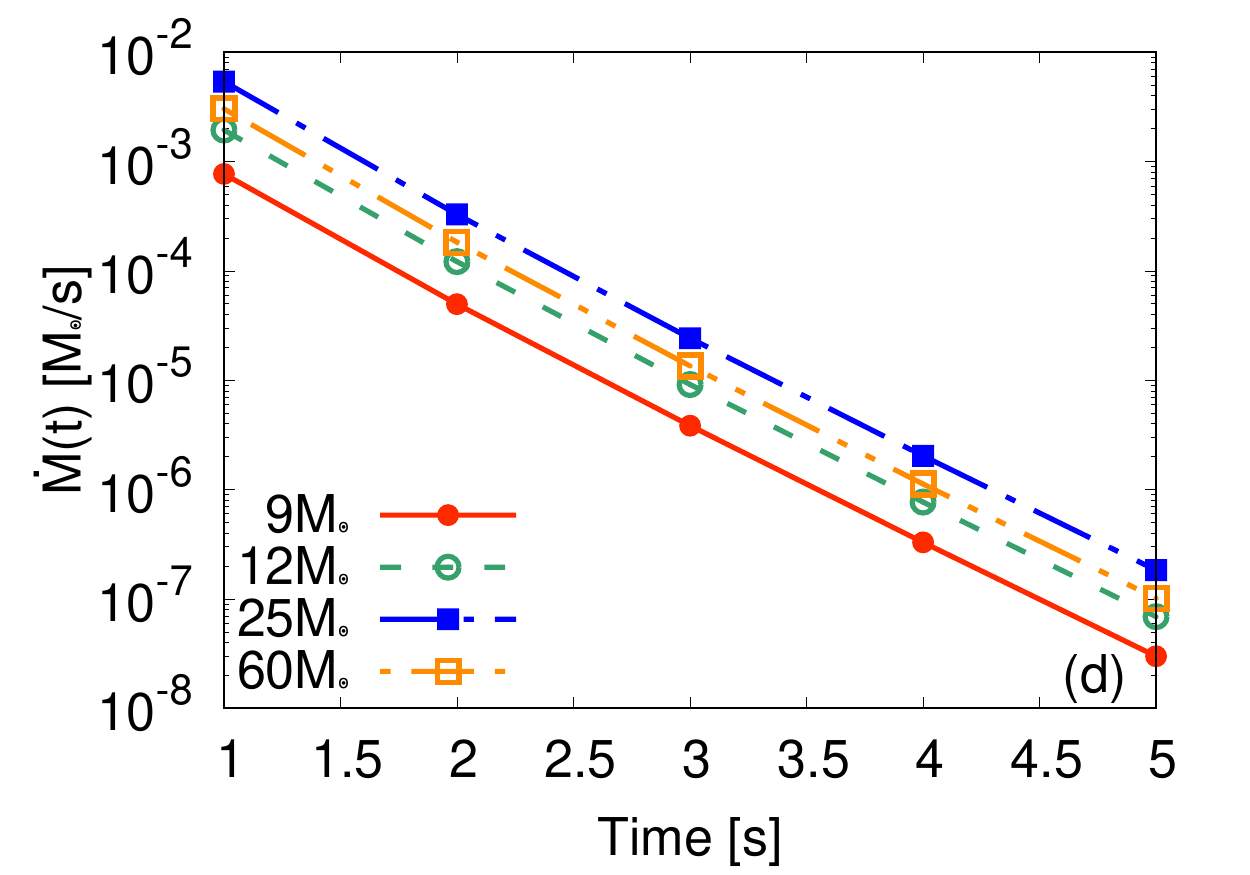}}%
\caption{Time evolution of hydrodynamic quantities relevant to neutrino driven winds including: (a) the entropy per baryon; (b) the expansion time scale; (c) the electron fraction; and (d) the mass ejection rate in different SN II progenitors.
}
\label{fig:hydro}
\end{figure}

The $\nu p$-process is sensitive to hydrodynamic quantities of neutrino-driven winds. Figure \ref{fig:hydro} shows the time evolution of various hydrodynamic quantities relevant to  neutrino driven winds such as the entropy per baryon $S$, the expansion time scale $\tau_{\mathrm{dyn}}$, the electron fraction $Y_{e}$, and the mass ejection rate $\dot{M}(t)$ in our SN II models. Here, we show the values of entropy and the expansion time scale at high temperature before the nucleosynthesis of heavy elements ($T=0.5$ MeV) as in \cite{Otsuki2000}. These hydrodynamic qunatities are sensitive to the mass of the PNS. The masses of PNSs in our SN II model are shown in Table \ref{tab:yields in different progenitor models}. As the mass of the PNS becomes large, $S$ ($\tau_{\mathrm{dyn}}$) becomes higher (lower) at each explosion time. In addition, the values of $S$ ($\tau_{\mathrm{dyn}}$) are increasing (decreasing) as the explosion time  passes. Such hydrodynamic properties of $S$ and $\tau_{\mathrm{dyn}}$ are consistent with results in \cite{Otsuki2000,Wanajo2013}. 
The nucleosynthesis inside neutrino-driven winds is sensitive to the electron fraction. The electron fractions $Y_{e}$ at the beginning of the network simulation ($T=10^{10}$ K) are shown in Figure \ref{fig:hydro}(c). The neutrino-driven winds are proton-rich ($Y_{e}>0.5$), so that the $\nu p$-process occurs in our wind model. The values of $Y_{e}$ are increasing irrespective of the progenitor model because the difference of $\average{{E}_{\bar{\nu}_{e}}}-\average{E_{\nu_{e}}}$ is smaller as the explosion time passes. As shown in Figure \ref{fig:hydro}(d), the mass ejection rates $\dot{M}(t)$ of the winds are smaller because of decreasing neutrino luminosities, so that the contribution of the later phase ($t>5$ s) to the total yields of $p$-nuclei is negligible in our wind model.

\begin{table*}[htb]
\begin{center}
\small
\begin{tabular}{|c c c|c c c c|} \hline
Type&Progenitor Mass&PNS Mass&$^{92}\mathrm{Mo}$&$^{94}\mathrm{Mo}$&$^{96}\mathrm{Ru}$&$^{98}\mathrm{Ru}$\\ \hline
SN II&$9$&$1.2$&$1.1\times10^{-15}$&$1.4\times10^{-16}$&$2.5\times10^{-17}$&$2.4\times10^{-18}$\\ 
SN II&$12$&$1.4$&$1.6\times10^{-14}$&$2.0\times10^{-15}$&$4.0\times10^{-16}$&$3.6\times10^{-17}$\\ 
SN II&$25$&$1.7$&$8.2\times10^{-12}$&$1.8\times10^{-12}$&$4.9\times10^{-13}$&$7.7\times10^{-14}$\\ 
SN II&$60$&$1.5$&$4.4\times10^{-13}$&$6.7\times10^{-14}$&$1.4\times10^{-14}$&$1.5\times10^{-15}$\\ 
HN&$100$&$3$&$2.0\times10^{-7}$&$1.5\times10^{-7}$&$1.6\times10^{-7}$&$2.9\times10^{-7}$\\ \hline
\end{tabular}
\end{center}
\small
\caption{PNS masses and yields of the $p$-nuclei in the $\nu p$-process of different progenitor models (in $M_{\odot}$).}\label{tab:yields in different progenitor models}
\end{table*}

\begin{figure}
    \centering
    \includegraphics[width=0.6\linewidth]{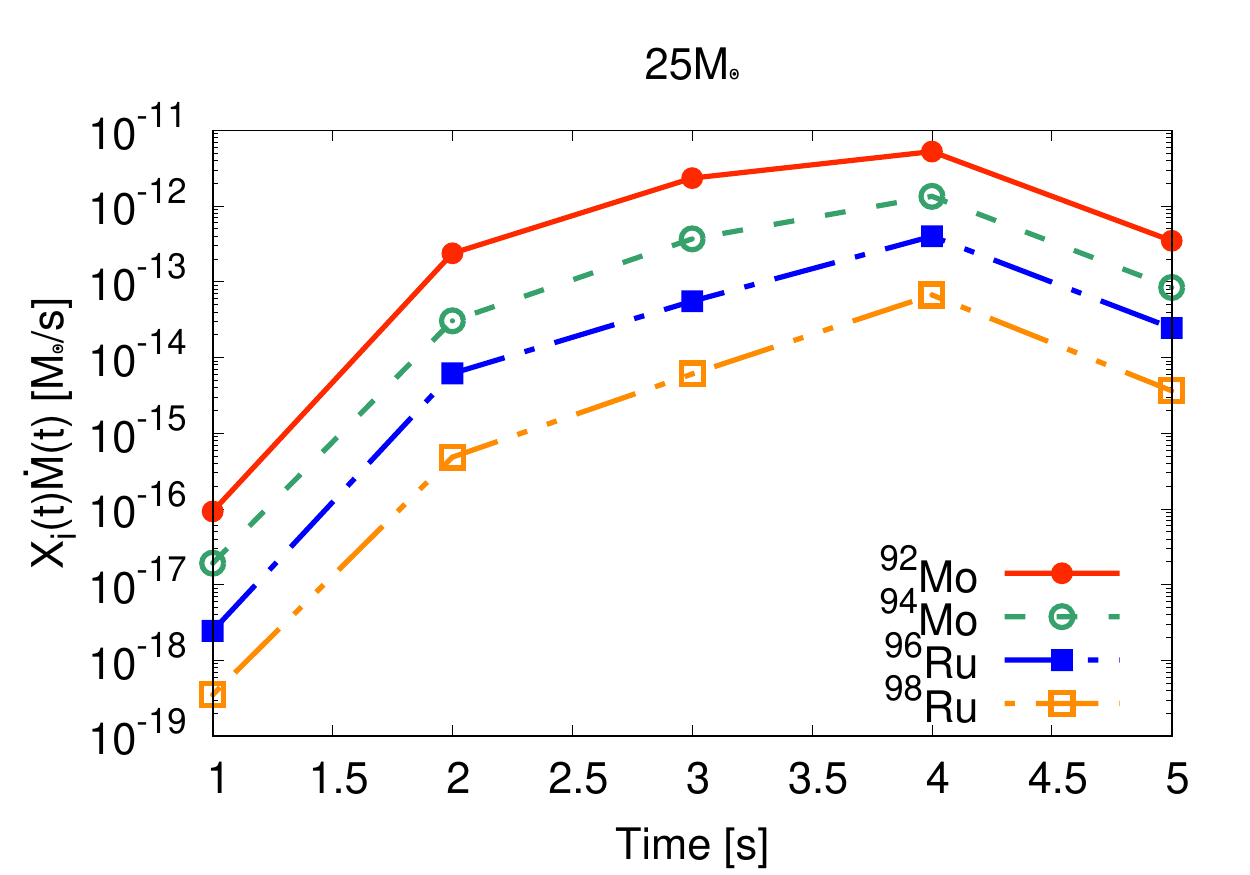}
    \caption{Time evolution of the $p$-nuclear yields in the $\nu p$-process of the $25M_{\odot}$ SN model.}
    \label{fig:vp25M}
\end{figure}

Figure \ref{fig:vp25M} illustrates the $\nu p$-process production of $p$-nuclei inside neutrino driven winds at different times during a SNe II explosion.  This figure shows the time evolution of the yields of $^{92,94}\mathrm{Mo}$ and $^{96,98}\mathrm{Ru}$ produced through the $\nu p$-process in the $25M_{\odot}$ SN II model. The total yield $y_{i}$ is given by the time integration of $X_{i}(t)\dot{M}(t)$ as defined above. 

General trends of the other progenitor models of $9M_{\odot},12M_{\odot}$, and $60M_{\odot}$ are similar although the value of $y_{i}$ depends upon the progenitor models. The $\nu p$-process is a primary process that proceeds without pre-existing seed nuclei originating from earlier generations of stars. 
The seed nuclei heavier than helium are synthesized in the early phase before the $\nu$p-process \citep{Wanajo2011} and the nucleosynthesis flow proceeds from these seed nuclei due to reactions with light particles such as protons and $\alpha$-particles. 

The production yields of the $\nu p$-process depend on the ratio of the number of free neutrons produced via $p(\bar{\nu}_{e},e^{+})n$ to the number of seed nuclei, denoted by $\Delta_{n}$ \citep{Nishimura2019}. Heavy $p$-nuclei are synthesized through the $\nu p$-process for  high $\Delta_{n}$. Becuase the values of $\Delta_{n}$ in the wind trajectories of the $25M_{\odot}$ SN II model are smaller than $10$, the $\nu p$-process does not effectively produce heavy nuclei around the mass number $A\simeq100{\--}110$ \citep{Wanajo2011}. Thus, the yields of the $p$-nuclei presented in Figure \ref{fig:vp25M} are larger than those of  $p$-nuclei heavier than $A = 100$ at any explosion time. The yield of each $p$-nucleus increases until $t=4$ s because the entropy per baryon $S$ also increases with decreasing neutrino luminosity (see Figure \ref{fig:hydro}(a)). Such production of heavier $p$-nuclei in a wind trajectory with the higher entropy is also confirmed in previous studies \citep{Wanajo2011,Nishimura2019}. 


In contrast, the smaller neutrino luminosity at $t>4$ s suppresses the production of $p$-nuclei. This is because the larger expansion time scales of the wind trajectories at $t>4$ s induce the production of large amounts of seed nuclei from which heavy elements are produced \citep{Otsuki2000,Wanajo2013,Xiong2020}; this reduces the value of $\Delta_{n}$ [e.g., see Eq.($9$) of \citet{Xiong2020}]. The mass ejection rate of each wind trajectory continues to decrease, and hence, the contribution from the later explosion phase ($t>5$ s) is negligibly small in the present SNe II models.

Table \ref{tab:yields in different progenitor models} shows the masses of the PNSs and the yields of $^{92,94}\mathrm{Mo}$ and $^{96,98}\mathrm{Ru}$ produced through the $\nu p$-process in the adopted progenitor models. The $9, 12, 25$, and $60M_{\odot}$ models correspond to the SNe II and the $100M_{\odot}$ model corresponds to the HN. The yields of $p$-nuclei are sensitive to the mass of the PNS. 
Figure \ref{fig:hydro} shows that in the case of massive PNSs, the entropy per baryon of the wind trajectories becomes large and, in contrast, the expansion time scales becomes small. Both conditions of the higher entropy and the shorter expansion time scale result in a higher value of $\Delta_{n}$ and a larger production of heavy $p$-nuclei in SN II progenitor models having a more massive PNS. The yields of $p$-nuclei in the HN model is much larger than those in the SNe II models because the PNS of the HN model ($M_{\mathrm{PNS}}=3M_{\odot}$) is much heavier than those of the SNe II models ($M_{\mathrm{PNS}}<2M_{\odot}$), which results in much shorter dynamical time scales in the HN wind \citep{Fujibayashi2015}. This implies that the $\nu p$-process in HNe can play an important role in the GCE of Mo and Ru.

The yields of $p$-nuclei produced through the $\nu p$-process in SNe II are much smaller than those in HNe.  We note, however,  that the potential effects of the reverse shock and fast neutrino flavor conversions \citep{Xiong2020} might enhance the yields of $p$-nuclei in SNe II. Such uncertainties of the $\nu p$-process are again discussed in Sec.\ref{sec:Conclusion}.


\section{GCE}
\label{sec:method}

We have performed  GCE calculations that  include contributions from the $\nu p$-process together with various other nucleosynthetic processes.   We have then analyzed the relative contribution of the $\nu p$-process to the observed solar and stellar abundances of $^{92,94}\mathrm{Mo}$ and $^{96,98}\mathrm{Ru}$.
 There are seven stable Mo isotopes. Among those, $^{92}$Mo and $^{94}$Mo are $p$-nuclei while  $^{96}$Mo is an $s$-only nucleus because it is shielded by the stable isobars $^{96}$Ru and $^{96}$Zr against $\beta^{\pm}$ decays. The other isotopes are produced by the $s$- and $r$-processes. The nucleosynthetic origin of Ru is similar to that of Mo. The isotopes $^{96}$Ru and $^{98}$Ru are $p$-nuclei and $^{100}$Ru is an $s$-only isotope, whereas the other Ru isotopes are synthesized by both the $s$- and $r$-processes. 
 
 The framework of our GCE calculations for $p$-nuclei is based upon the one-zone model of \citet{timmes1995,yamazaki2021}. 
This model includes  exponentially decaying galactic inflow and a star formation rate  calculated using a  quadratic Schmidt function \citep{larson1969}.

The evolution of $s$- and $r$-nuclei with metallicity is taken from \citet{kobayashi2020} and \citet{yamazaki2021}, respectively, except for $p$-nuclei. Although the GCE model used in \citet{kobayashi2020} is different from ours, 
the GCE of the $s$-process depends very weakly on adopted models and the result is rather robust. 
Moreover, because the $s$-process is a secondary process and its progenitor is a long-lived low- or intermediate-mass star, it does not  contribute strongly to the early GCE. 

The production rate of each $p$-nucleus is derived from  the event rate of SNe (including HNe) and the ejected mass of the synthesized $p$-nucleus associated with each progenitor.
The mass range of HNe constrains its event rate through the initial mass function (IMF). We adopt the Kroupa IMF \citep{kroupa2001} and set the mass range of zero-age main-sequence (ZAMS) stars for SNe and HNe to be equal to 8$\--$60$M_\sun$ and 60$\--$100$M_\sun$, respectively. In this configuration, $4\%$ of massive stars explode as HNe and this fraction is consistent with the value recently deduced  in \citet{shivvers2017}.
The delay time $\tau$ of the SN (HN) explosion of a massive star due to stellar evolution is set to be equal to the main sequence lifetime as a function of the ZAMS mass.


The input data of the $\nu p$-process in the different progenitor models are summarized in Table \ref{tab:yields in different progenitor models}. There are many controversial results published in the literature regarding the $\gamma$-process yields. The reader should be aware that the $\gamma$-process yields plotted in Figure 3 and 4 were obtained from the calculations of \citet{Kusakabe2011} for SNe Ia and from \cite{Travaglio2018} for SNe II. It has been  pointed out \citep{Travaglio2015} that the $\gamma$-process in SNe Ia contributes to the solar abundances of the $p$-nuclei. In our GCE calculation of  $p$-nuclei, we take into account the contributions from SNe Ia by using the yields of $p$-nuclei from \citet{Kusakabe2011} in Case A1. Another possible major process is the $\gamma$-process in SNe II. To include the production of the SN $\gamma$-process in GCE, we employ the yields from the {\sl {\it xi}45} series of the KEPLER model \citep{Travaglio2018} where the contribution from the nucleosynthesis inside neutrino-driven winds is not included. The yield data of the {\sl {\it xi}45} model are given for seven different initial progenitor masses from $M=13M_{\odot}$ to $30M_{\odot}$ and $15$ different metal mass fractions from $Z=0$ to $3.1\times10^{-2}$.

We note that the SNe II $\gamma$-process is caused by photodisintegration in the outer O-Ne-Mg layer at $^>_\sim$10000 km). On the other hand, the $\nu p$-process occurs inside the neutrino-driven winds near the atmosphere of the PNS at $\sim$100 km. It is, therefore,  clear that the contribution from the $\nu p$-process in the neutrino-driven winds is not included in the {\sl {\it xi}45} KEPLER models at all. The {\sl {\it xi}45} KEPLER model provides nuclear yields of $p$-nuclei produced through the SNe II $\gamma$-process in different progenitor masses and metallicities. Table \ref{tab:yields in different progenitor models} is the only input data of the $\nu p$-process yields in our GCE calculations. We note that more robust and quantitative conclusions could be obtained if we used much larger numbers of progenitor masses and metallicities for the $\nu p$-process calculations. However, as far as we know, there is no publication of the yield data from the $\nu p$-process in HNe having more massive PNSs except for that of \cite{Fujibayashi2015}. It is beyond the scope of the present study to analyze the progenitor dependence of the $\nu p$-process in  HNe.

\begin{figure}
    \centering
    \includegraphics[width=0.6\linewidth]{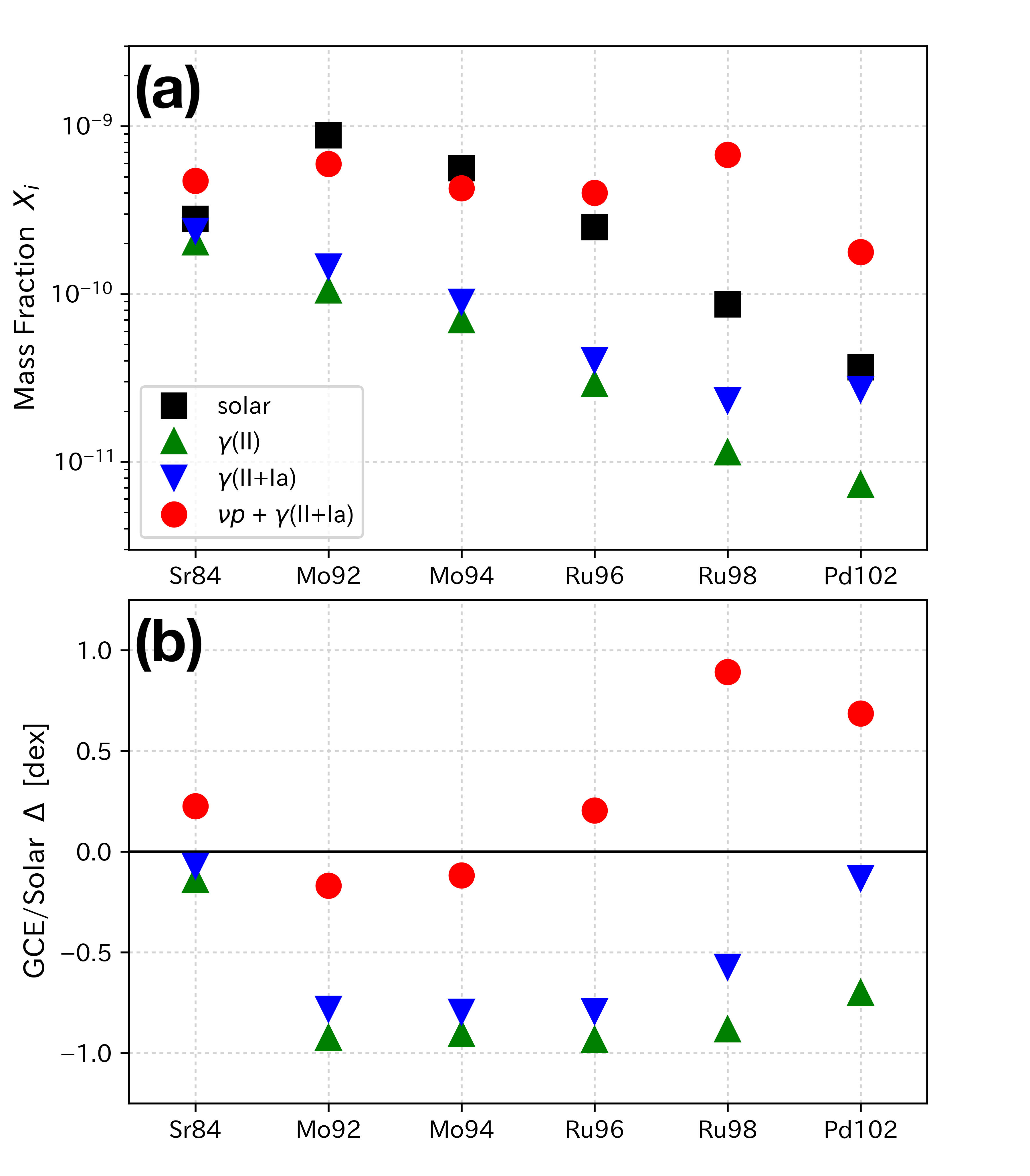}
    \caption{Final abundances of the $p$-nuclei at [Fe/H]=0 compared with the solar system abundances.}
    \label{fig:}
\end{figure}

Figure~\ref{fig:}(a) shows the final mass fractions of the $p$-nuclei at solar metallicity derived from the GCE calculation (circles).  These are compared with the solar abundances (squares). The up and down triangles correspond to the contributions from the $\gamma$-process in SNe II only  or the sum of SNe II and Ia, respectively. The red circles correspond to the total yields from the  $\nu p$-process plus $\gamma$-process. Figure \ref{fig:}(b) shows the ratios of the theoretical to observed solar abundances on a  logarithmic scale. The contribution from the SN $\gamma$-process alone underproduces the solar abundances of $^{92,94}$Mo and $^{96,98}$Ru, and the contribution from SNe Ia is insignificant. In contrast, the $\nu p$-process significantly enhances the theoretical abundances of these $p$-nuclei. The contribution from the $\nu p$-process in Figure \ref{fig:} mainly comes from HNe as shown in Table \ref{tab:yields in different progenitor models}. 
This smaller effect of the $\nu p$-process in SN II is consistent with implications in \citep{Xiong2020,Jin2020} in the case without a reverse shock and neutrino fast flavor conversions. A HN (SN) progenitor having a massive PNS enhances the contribution of the $\nu p$-process to the solar abundances of the $p$-nuclei. The overproduction of $^{98}\mathrm{Ru}$ and $^{102}\mathrm{Pd}$ caused by the $\nu p$-process reflects the dominant production factors in our HN wind trajectory [see the middle panel of Figure $12$ in \citet{Fujibayashi2015}]. 
This overproduction of heavy $p$-isotopes, however,  may be suppressed if we used a recently proposed large triple-$\alpha$ reaction rate \citep{Jin2020} in our network calculations because their proposed four-body reaction mechanism may enhance the abundance of seed nuclei.

\begin{figure}
\subfigure{%
    \includegraphics[clip, width=0.5\columnwidth]{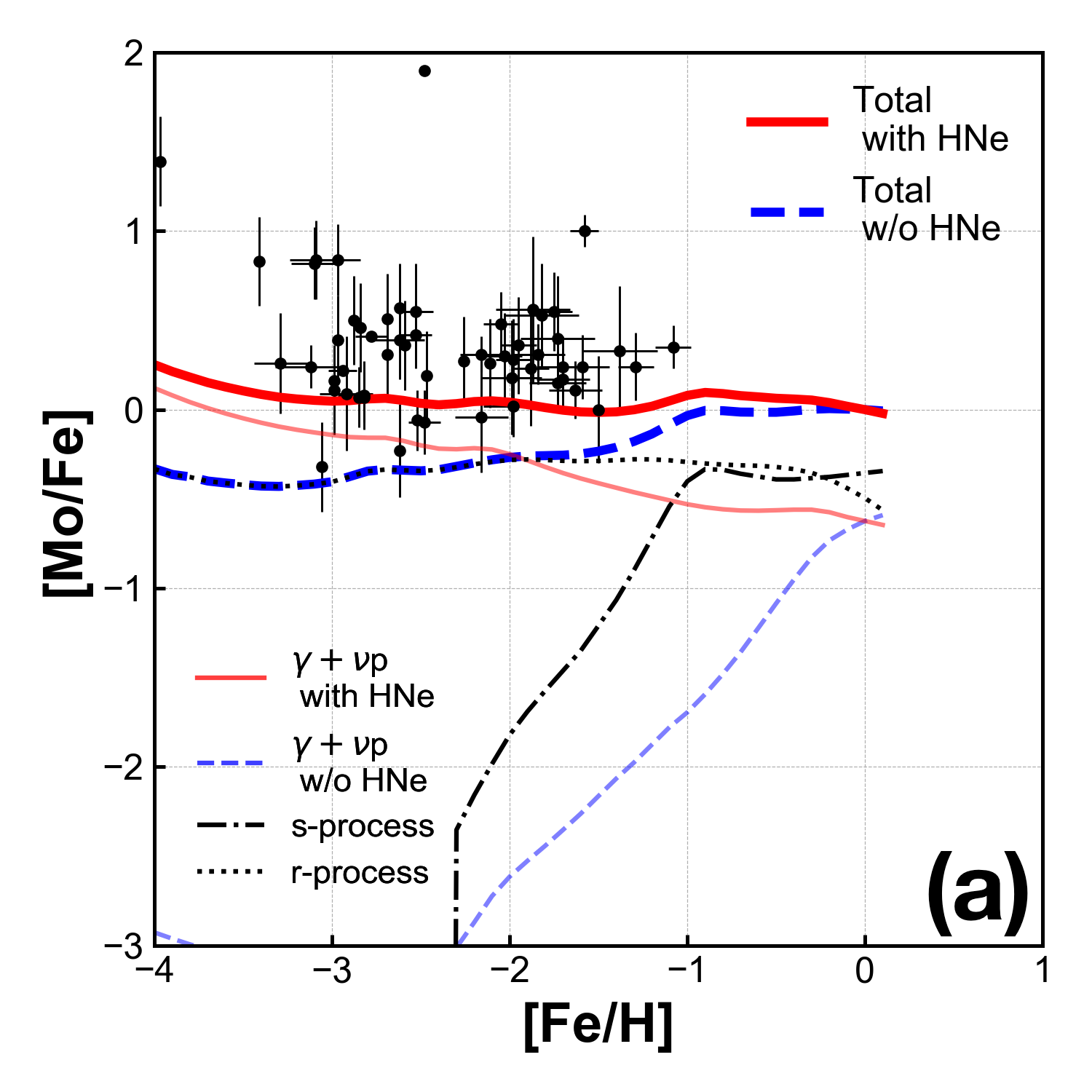}}%
\subfigure{%
    \includegraphics[clip, width=0.5\columnwidth]{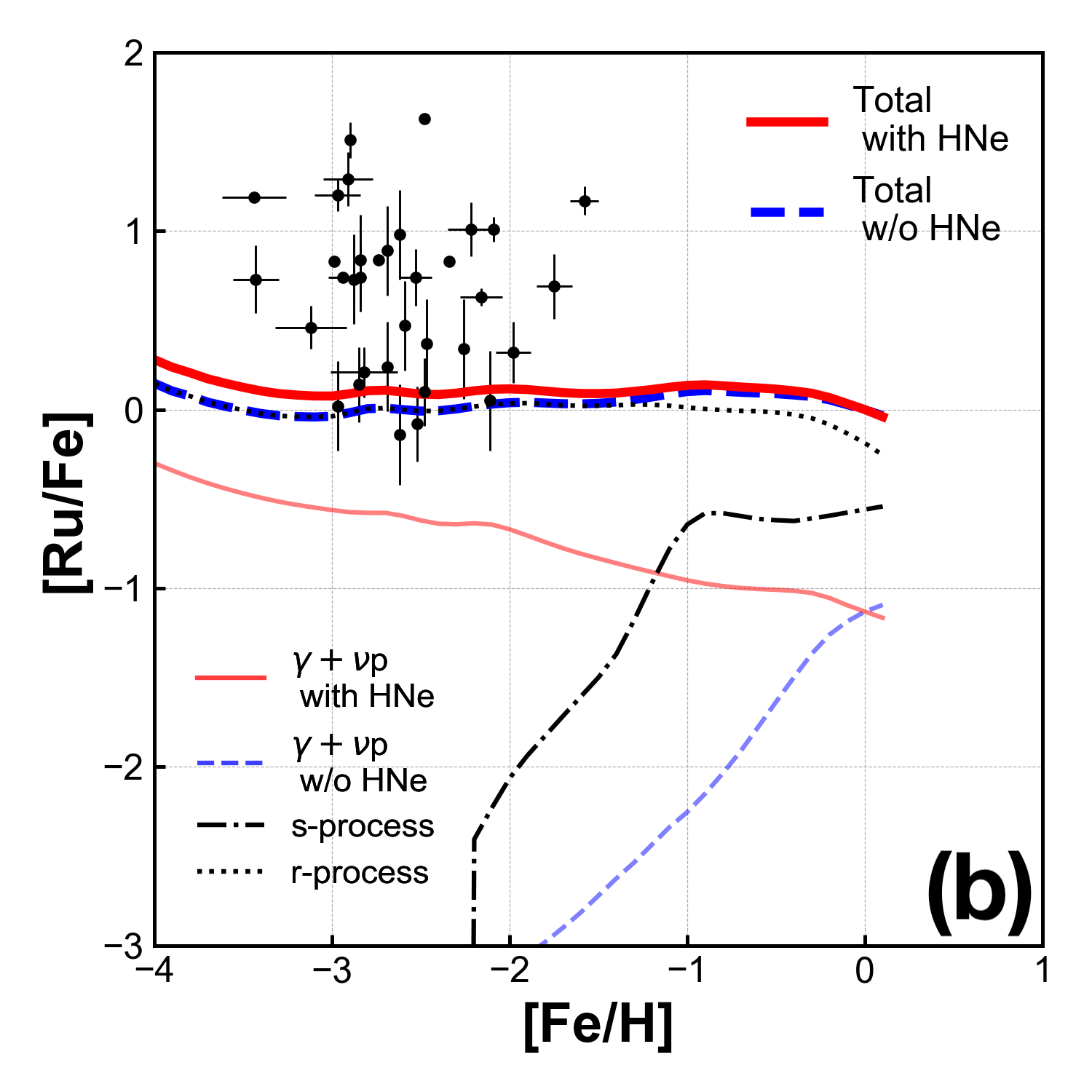}}%
\caption{Elemental abundance evolution of (a)Mo and (b)Ru. Thin lines show each process contribution. Thick lines are the total elemental abundances. Solid lines include the contribution of the $\nu p$-process in HNe while the dashed lines do not.}
\label{fig:gce}
\end{figure}

\begin{figure}
\subfigure{%
    \includegraphics[clip, width=0.5\columnwidth]{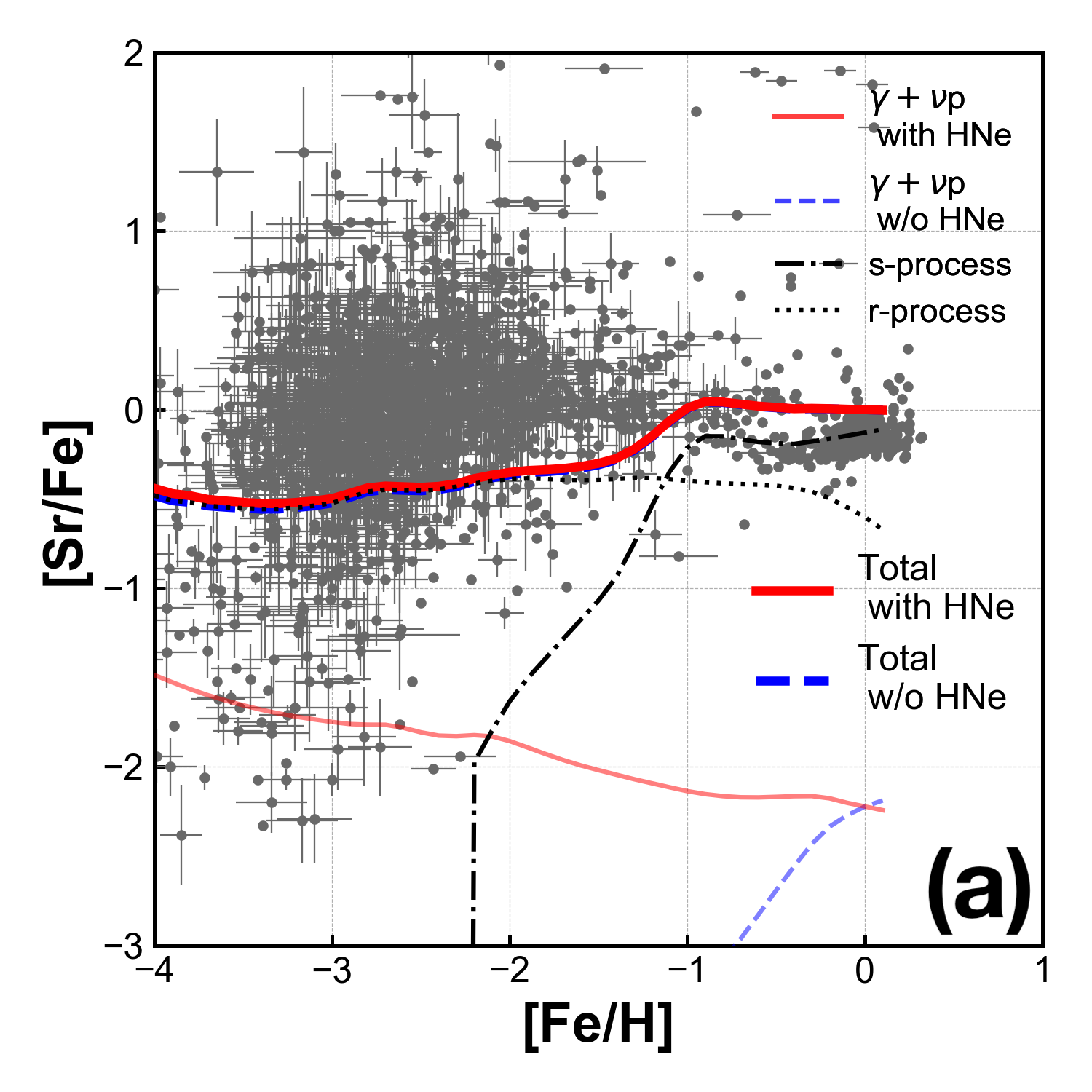}}%
\subfigure{%
    \includegraphics[clip, width=0.5\columnwidth]{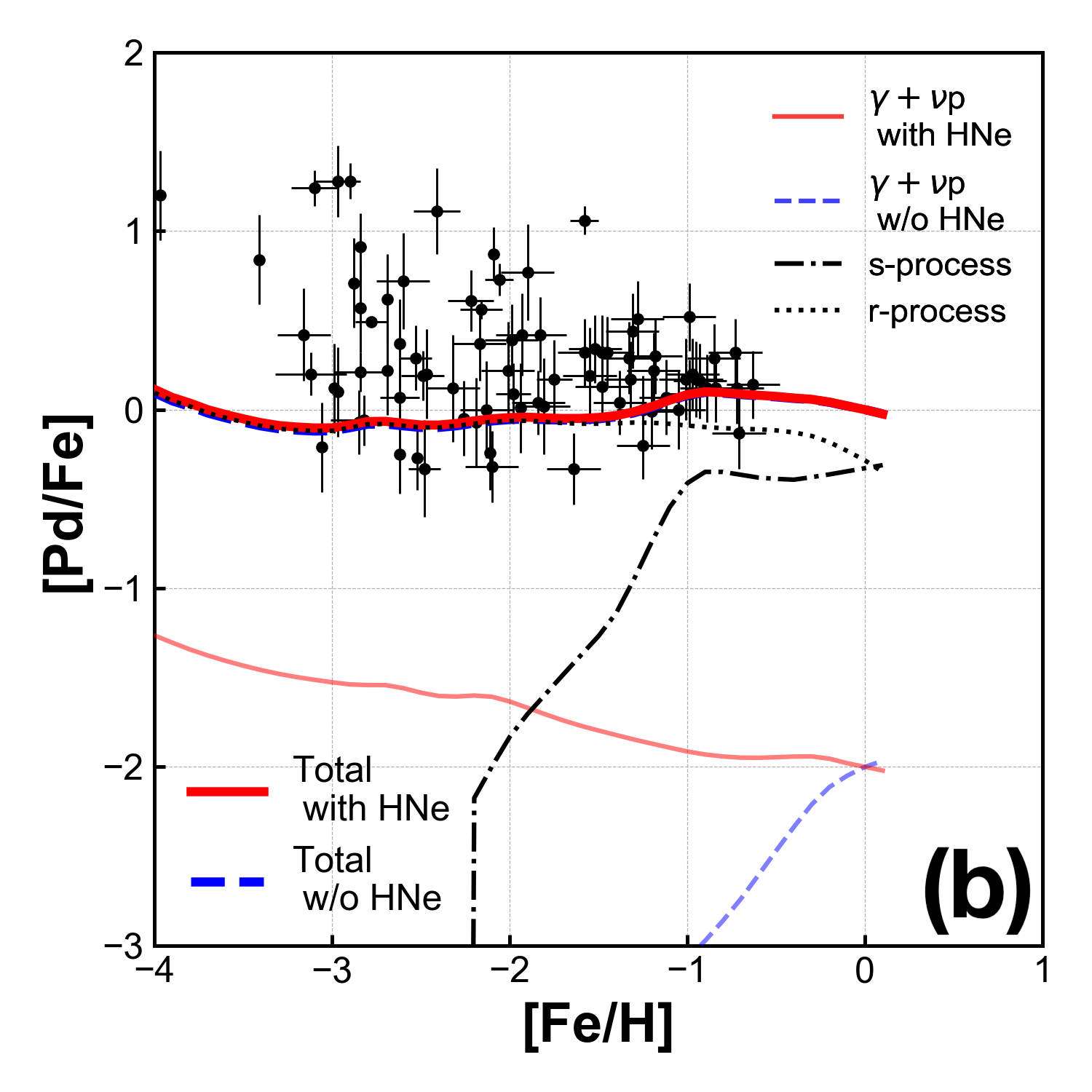}}%
\caption{Elemental abundance evolution of (a) Sr and (b) Pd as in Fig.\ref{fig:gce}.}
\label{fig:gce_Sr_Pd}
\end{figure}


Figure \ref{fig:gce} displays the elemental number ratios [X/Fe] against [Fe/H], which are normalized at [Fe/H]=0.
The filled circles are the observed stellar data taken from \textcolor{red}{}the SAGA database \citep{sagadatabase}.
The black dotted and dashed-dotted lines show the abundances from the $r$-process and the $s$-process, respectively.
The thin dashed (blue) line in Figure~\ref{fig:gce}(a) shows the summed abundances of the $p$-isotopes ([($^{92}\mathrm{Mo}+^{94}\mathrm{Mo})/$Fe]) produced through the $\gamma$-process and the $\nu p$-process in SNe without the $\nu p$-process from HNe.
The thin solid (red) line  shows the abundances of the Mo $p$-isotopes produced from the $\gamma$-process, the $\nu p$-process in SNe, and the $\nu p$-process in HNe.
The thick solid (red) line  and the thick dashed (blue) line show the total elemental abundances in the present GCE model calculation with and without the contribution from the $\nu p$-process in HNe, respectively. 
As shown by the thin dashed  (blue) line, the $\gamma$-process and $\nu p$-process from SNe alone do not significantly account for the evolution of the Mo abundance over the entire metallicity range.
In the region of [Fe/H] $<$ -1, the contribution of the $s$-process is also negligibly small.

In contrast, the $r$-process (dotted line) makes a relatively large contribution for [Fe/H] $<$ -1, but the amount of the $r$-nuclei is somewhat lower than the observed stellar ratios.
As shown in the thick and thin solid lines (red), the $\nu p$-process in HNe significantly enhances the abundance of the $p$-isotopes and, in the region of [Fe/H] $<-2$, the production by the HN $\nu p$-process is larger than that of the $r$-process.
Thus,  HNe are the dominant contributor to the  $\nu p$-process for [Fe/H]$<-2$.
The present result also indicates that the observed Mo abundances in the low metallicity region are predominantly produced through the $\nu p$-process in HNe. 
Because population III stars in the early Galaxy are thought to be massive $\sim 100M_{\odot}$ \citep{Hirano2014}, our conclusion of the HN dominance in the low metallicity region seems reasonable.

Figure~\ref{fig:gce}(b) similarly shows the elemental abundance of Ru with and without the $\nu p$-process in HNe.  The $\gamma$-process and the $\nu p$-process in SNe II hardly affect the elemental abundances of Ru, and the $\nu p$-process in HNe increases the elemental abundance in the low metallicity region.
The contribution from the HN $\nu p$-process on the Ru elemental abundance is not as prominent as for Mo. 
This is because the total solar isotopic fraction of the $p$-isotopes $^{92,94}\mathrm{Mo}$ is as high as $24.1$\% but that of Ru ($^{96,98}\mathrm{Ru}$) is only $7.4$\%.
We note that $^{92,94}\mathrm{Mo}$ could also be synthesized in slightly neutron-rich ejecta ($Y_{e}\sim0.47$) \citep{Bliss2018,Wanajo2018}.
If an early neutron-rich ejecta of core-collapse SNe at $t<1$s were  taken into account in the GCE calculations, the total abundances of Mo and Ru could increase and the theoretical prediction become more consistent with the observational stellar abundances.

Figure.\ref{fig:gce_Sr_Pd} illustrates  how our GCE model describes the evolution of other  
elements. This figure  displays the  mass  ratios  [Sr/Fe] and [Pd/Fe]  against  [Fe/H]. We chose these two elements because their abundances in the solar system are known to be dominated by the $s$-process and $r$-process contributions~\citep{bist14}.
In these nuclei the pure $p$-process isotope is limited to $^{84}$Sr (isotopic fraction is 0.56\%) and $^{102}$Pd (1.02\%). 

The contribution from the $\nu p$-process in HNe increases as the metallicity becomes lower (red thin solid curves) in Figure \ref{fig:gce_Sr_Pd} similarly to Mo and Ru in Figure.~\ref{fig:gce}. However, both Sr and Pd exhibit differences from the GCE of Mo and Ru such that the s+r components dominate in the entire  metallicity range of $-4 <$ [Fe/H] $< 0$. The contribution from the $\nu p$- and $\gamma$-processes is less than $1\%$ even at [Fe/H] $= -4$ in the both cases with and without contributions from HNe.
We therefore emphasize the significance of Mo and Ru in particular to study the GCE, for their substantial pure $p$-isotopic fractions.

Our GCE model 
provides the abundance curves that pass through the bottom of observed elemental abundance distributions in Figures.~\ref{fig:gce} and ~\ref{fig:gce_Sr_Pd}. The observed abundance scatter above our theoretical curves is mainly due to the inhomogeneous nature of SN and HN ejecta including explosive nucleosynthetic products~\citep{ishimaru99,WanajoIshimaru2006}
in the metal-poor epoch of GCE.  This applies to all four elemental abundance distributions for Mo, Ru, Sr and Pd.

 The GCE of the $\nu p$-process elements may more or less depend on various quantities such as the initial mass of progenitor stars, metallicities, SN or HN event rate, and so on. 
It is highly challenging to take into account all of these quantities. Here, we estimate the sensitivity to 
the progenitor mass range of HNe by adopting
three cases: $60-120 M_{\odot}$; $60-100 M_{\odot}$; and $40-100 M_{\odot}$, keeping the same nucleosynthetic yields of HNe.
We then find a difference of less than $0.2$ dex  in [X/Fe] which is only apparent in the lower metallicity region [Fe/H] $< -2.5$ among the three cases. 
Therefore, our GCE result is not sensitive to the progenitor mass range of HNe.

As was discussed previously, the enhancement of the isotopic abundances of $^{92,94}$Mo correlates with that of the $r$-isotope $^{100}$Mo in primitive meteorites \citep{Dauphas2002,Budde2016,Poole2017}.
Core-collapse SNe are still a viable candidate for the astrophysical site of the $r$-process. If a single SN happened near the proto-solar nebula and both the $\nu p$- and $r$-processes occurred in the SN, their products would have contaminated the proto-solar material as observed in primitive meteorites.

\section{Conclusion}
\label{sec:Conclusion}
We have calculated the $\nu p$-process contribution in the neutrino-driven winds of core-collapse SNe (in particular including HNe) to the GCE of $^{92,94}\mathrm{Mo}$ and $^{96,98}\mathrm{Ru}$.
We have shown that the contribution of the $\nu p$-process in ordinary SN II is negligibly small, while  that of the HN $\nu p$-process 
with massive PNSs dominates. 
The HN $\nu p$-process contribution to the GCE of the $p$-nuclei is largest at low metallicity.
The high [Mo/Fe] ratios observed in metal-poor stars indicate that the $\nu p$-process in HNe is the dominant source of $^{92,94}\mathrm{Mo}$ and $^{96,98}\mathrm{Ru}$ in the Galaxy.

This study has explored  the possible effect of the $\nu p$-process on the GCE of $^{92,94}\mathrm{Mo}$ and $^{96,98}\mathrm{Ru}$.  We note, however, that there are uncertainties in this analysis. First, we have not considered the uncertainties due to reverse shock effects and fast flavor conversions of neutrinos. The potential impact of the reverse shock and neutrino fast flavor conversions was discussed in \cite{Xiong2020}. Although these might enhance the $\nu p$-process in SN II and induce a non-negligible influence on the GCE of Mo and Ru, these effects are still subject to large theoretical uncertainties. Long-term simulations of core-collapse supernovae beyond $1$s are required to more precisely include such uncertainties in the $\nu p$-process. A second caveat is that  we estimate yields of $p$-nuclei based upon one specific HN model and the dependence of these results on progenitor mass and metallicity in HN is not taken into account in our GCE calculations. It seems likely  that GCE results of $p$-nuclei are not particularly  sensitive to the progenitor mass range of HN.  Nevertheless,  further studies of the progenitor dependence of HN are required to obtain a  more quantitative conclusion.

\section*{Acknowledgement}
We thank Alexander Heger for providing data of {\sl {\it xi}45} in KEPLER model. This work was supported in part by Grants-in-Aid for Scientific Research of JSPS (19J13632, 20K03958, 21J11453). Work at the University of Notre Dame (GJM) supported by DOE nuclear theory grant DE-FG02-95-ER40934. Work at Soongsil University  was supported by the National Research Foundation of Korea (Grant Nos. NRF-2020R1A2C3006177 and NRF-2013M7A1A1075764)

\bibliography{ref_sasaki,ref_YY}{}
\bibliographystyle{aasjournal}



\end{document}